\documentclass[12pt]{article}
\usepackage{graphicx}

\begin{document}
\title{Magic Numbers from New Systematics}
\author{Chinmay Basu\\
Saha Institute of Nuclear Physics ,\\ 1/AF Bidhan Nagar,
Calcutta 700064, India}
\date{}
\maketitle
\begin{abstract}
A new systematics from the separation energy of deuteron is used to examine the
magicity of stable as well as nuclei towards dripline.  
\end{abstract}

\newpage
Nuclei with N (neutron mumbers)=2, 8, 20, 28, 50, 82, 126 show unusual stability 
and hence are called magic nuclei. Similarly Z=2, 8,20, 28, 50, 82 are also 
established proton magic numbers. Whether these magicities change for proton or
neutron rich nuclei has been a subject of recent interest [1-4]. 
In course of these works, the well known magicities have been shown
to loose their character with inreased neutron or proton numbers 
and some new magic numbers have evolved. These attempts 
use different systematics;1) the plot of one and two neutron or proton
separation energies, 2) the Q-value for $\beta^-$ and $\beta^{+}$ decay
3)excitation energy of the 1st excited state and 4) the energy of alpha
particle for examining the magicity of nuclei. 
There are also indirect methods such as calculating the shell gap
from two nucleon separation energy. 
The different systematics have normally agreed in case of well known magic 
numbers but there are some
differences when one comes to the cases of not so well known magic numbers. 
For example in Ref[4] $N=16$ has been shown to be a magic number for 
$T_z=3, 5/2, 7/2 (Z=9,10,11)$ from breaks in 1-neutron separation 
energy plotted against
odd neutron numbers for a given $T_z$. 
This magic number is later confirmed from $Q_{\beta_{-}}$ systematics 
for $T_z=5/2,7/2$ only [3]. In ref[1] on the other hand the authors 
plot the 1-neutron
separation energy against all N (both even and odd) at a given Z 
(both odd and even). They conclude from 
comparison with a model that $N=16$ magicity appears only at $Z=7,8$ and not
at Z=10 ($T_z=3$) or $Z=9,11$. The authors of [1-3]
also discuss a number of new magic numbers towards the dripline from
their respective systematics.

In this paper we present a new sytematics to investigate the indication
of magic numbers. The separation energy of deuteron ($S_d$) is plotted against 
nucleon numbers and the breaks in the systematic trend indicate magicity. 
No nuclei in it's ground state have been observed to decay by deuteron emission
(Neither do any nuclei show negative deuteron separation energy). 
Recent calculations also do not support nuclei having minimum energy 
configuration with a deuteron [5]. 
However, the ground state of the nucleus $^6$Li for example 
is well described by the $d+\alpha$ cluster structure [6]. 
Experimental studies of $\alpha$ knockout
reactions [7] and  back angle
increase of $\alpha$ elastic scattering cross-section from $^6$Li [8] also 
support it's deuteron
substructure. In Ref. [8], the possiblity of  
$^6$Li (in which the deuteron is loosely bound and it's separation energy 
is smaller than neutron
and proton separation energy) as a deuteron-halo nucleus has  
been discussed. 
The separation energy plots (in which the odd-even staggering in mass plots
is doubled) of deuteron have also been utilised earlier to exhibit the np
pairing in light odd-odd nuclei [9]. This motivates us to use the plot of
deuteron separation energy to investigate magicities near and away from the
stability line.
For traditional magic numbers we see that the breaks in $S_d$
 are rather striking
and are easily indicative. For not so striking indications we follow the
following prescription: If the separation energy of deuteron at any odd neutron
number say ($N$) falls below the line joinning the $N-2$ and $N-4$
nuclei then the isotope with $N-1$ is magic or extra stable. If 
on the other hand $S_d$ is above the line, magicity
is quenched. This prescription is also applied to indicate proton magicities.

The separation energy is defined as 
\begin{equation}
S_{d}(N,Z)=B(N,Z)-B(N-1,Z-1)-B_d
\end{equation}
where the binding energy of two nuclei and deuteron are involved. 
In fig. 1 we show the deuteron separation energy 
as a function of N at specific Z values. We choose Z values in such a way 
so as to scan the well known neutron magicities. We do not include in our plots
the systematics predicted values and show only the experimental data points.
All the experimental masses required for our calculations are adopted from
the latest mass table of Audi and Wapstra [10]. 
The very well known magicities of N=2(Z=4),
N=8(Z=8), N=20,28(Z=20), N=50,82(Z=50) and N=126(Z=82) are clearly indicated
by the new systematics. The 
separation energy has a increasing trend when plotted against neutron
numbers unlike the
decreasing trend of the 1n-separation energies.  
This makes it easier to observe 
the extra downward shift of $S_d$ at $N=N_M+1$ (where $N_M$ is a neutron 
magic number).
For Z=30 since 
no magic neutrons numbers are encountered, the systematics do not show 
any break. 
In fig. 2 we show the well known proton magicities from the 
plot of separation 
energy against proton numbers at fixed neutron numbers. 
The Z=8(N=8,16),Z=20(N=24),Z=50(N=65),
Z=82(N=102,126)  
are clearly visible from
the new systematics.
The loss of some of the well known neutron and proton magicities with neutron
or proton numbers are also studied. The loss of N=8
magicity at Z=4 could be seen from fig.1. 
 According to the $Q_{\beta^-}$ systematics[3] N=20
vanishes at $T_Z\geq{7/2}$($Z\leq{13}$) and according to [1] the
Z region is $Z=11-14,18,22-24$. In this work we find that N=20 magicity 
vanishes at Z=10-12 only and is present at Z=13 (fig.3),14 (fig.4). 
In the Z=22-24 region we could not conclude
about the behaviour of N=20 magicity due to unavailable data points in the
literature.
Disagreement with [1] is found for Z=18 (fig.3) 
where no quenching was observed for the N=20 magic number.
The N=28 magicity is shown to loose its character at $T_Z\geq{5}(Z\leq{18})$
from the $Q_{\beta^{-}}$ systematics [3] and at Z=17,25,26,30 from [1]. 
We find that this magicity is absent 
at Z=16(fig.4),17,18(fig.3) and 26(fig.3). It is present however at
Z=25 (fig.4). For Z=30 and Z$<$16, N=28 magicity could not be examined without
systematics predicted value.

The Z=8 magicity is observed from $Q_{\beta ^{-}}$ systematics in the region
N=11,13 whereas [1] predicts a loss of magicity at N=11 in the range
N=6,10-12. We find (fig.3) that the Z=8 proton magic number is quenched at N=11 thereby
disagreeing with [3].
The vanishing of Z=20 magicity at N=16-18,22-26 is reported in [1]. This agrees
with our systematics for N=16 (fig.2) and N=18 (fig.4). In the N=22-26 region 
the present systematics agrees with [1] for N=22,23 only
(N=23 in fig.3) where the Z=20 magicity is quenched.  
For the region N=24-26 the magicity though visible is fairly weak (e.g
N=24 (fig.2)).

Finally we investigate some of the newly discovered magic numbers in the 
framework of the present systematics. For example we observe the N=6
magic number at Z=4 (fig.1) in the region Z=3-8 predicted by [1].
Referring to fig.4 we show the N=16
magicity for Z=9,10 which was predicted to be present in 
$T_Z\geq{3}(Z\leq{10})$ region by Ozawa et al [4]. 
The N=30 magicity predicted
by [3] is observed in this work at Z=26 (fig.4) only. 
For the Z=14 magicity quite different N ranges are suggested in literature.
In [3] it is $21\leq{N}\leq{27}$ and in [2] it is $13\leq{N}\leq{19}$. 
From the present systematics we observe this magic number at N=16 (fig.2),
18(fig. 4)
which is in the range predicted by [2].  
The new magic number at Z=16  predicted in the range N=21-27 by [3] is observed 
from the new systematics at N=22-27(fig.3) but is found to be absent at 
N=21. Some of the discrepancies of different magic numbers and their behaviour
away from the stability line are summarised in Table 1. This shows the
importance of deuteron separation energy systematics in the proper
prediction of magic numbers and their behaviour. Therefore,in order to 
obtain a complete picture the present systematics along with other 
systematics must be considered.

In addition to the already observed magicities by the new systematics we
observe a magic like behaviour of N=26 at Z=13 and 14 (fig.3).  We also see
indication of extra stability for N=26 in P,S (fig 4) and Cl. 
A pseudo shell closure
at N=26 is reported for Cl,S,P isotopes 
from mass measurements at GANIL and Dubna [11]. 
The break in $S_2n$ versus N plot followed by a
flattening at N+2 and beyond indicates that  N is a magic number. 
This break was observed in [11]
with the measured mass data for Cl,S,P. 
However this data confirms the N=26 magicity only for Z=16,17 from the
flattening at N=28 in the $S_{2n}$ against N plot.
The present systematics 
clearly show the N=26 extrastability for Z=13-17 with the available mass data. 
Detailed investigation in this direction may be pursued in future. 

In this work we present a new systematics from a plot of deuteron 
separation energy against number of neutrons or protons to investigate 
the magicity of nuclei. The new
systematics reproduces the well known magic numbers. The behaviour of well
known magic numbers have been discussed with variation of neutron and proton
numbers and in comparison to other established systematics. The newly observed
neutron and proton magicities have also been discussed. The present
systematics indicates extrability of N=26 for Al to Cl isotopes.

\newpage
{\bf Table~1}  Some discrepancy in the behaviour of magic numbers as observed
from the present work in comparison to other publieshed works. The 
abrreviations(l) stands for lost, (q) for quenched, (p) for present and 
(c.s) cannot say.\\

\begin{center}
\begin{tabular}{cccc} \hline
{$\bf Magicity$}&${\bf Ref[1]}$&${\bf Ref[3]}$&${\bf This work}$\\ \hline 
N=20(l)&Z=11-14,18,22-24&Z$\leq$ 13&Z=10-12\\
&&&(Z=13,14,18,22,23(p), Z=24 (cs))\\
N=28(l)&Z=17,25,26,30&Z$\leq$ 18 &Z=16-18,26\\
&&&(Z=25,30 (p))\\
Z=8&N=6,10-12(l)&N=11,13(p)&N=12(p),N=10,11,13(l)\\
Z=20(l)&N=16-18,22-26&-&N=16,18,22,23\\
&&&(N=24-26 magicity weakly present)\\
N=6&Z=3-8(Ref.[2])&-&Z=3,4,6\\
&&&(Z=5(q),Z=8(c.s))\\
N=16&Z=7,8(Ref.[2])&Z=9-11&Z=7-11\\
N=30&-&Z=21,23&Z=26\\
Z=14&N=13-19(Ref[2])&N=21-27&N=16,18,25\\
&&&(N=21-24(l),N=26,27(c.s)\\
Z=16&-&N=21-27&N=22-27\\
&&&(N=21(l))\\
\hline
\end{tabular}
\end{center}

\newpage
\oddsidemargin=-0.2in

\begin{figure*}
\includegraphics{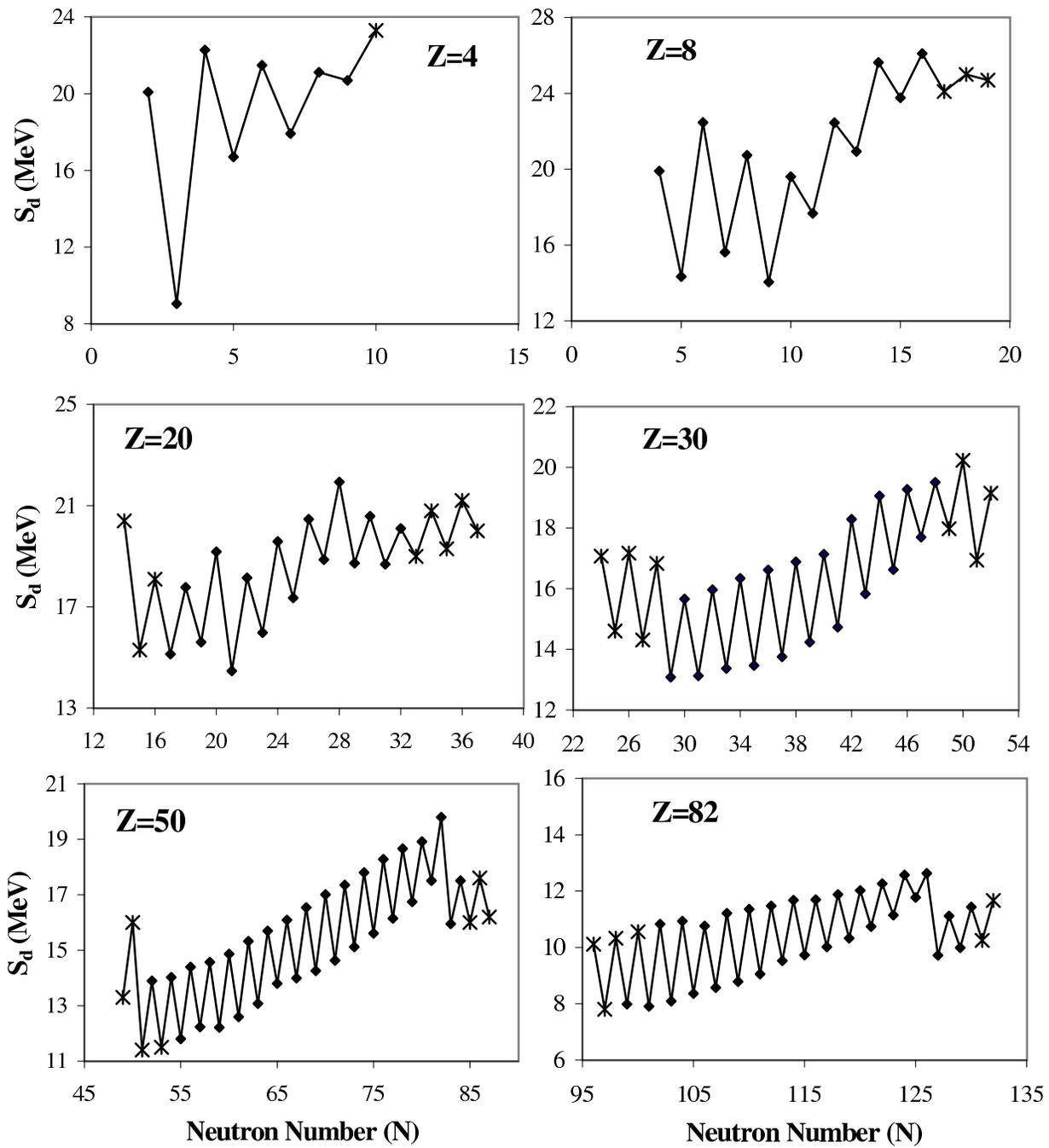}
\caption{Plot of deuteron separation energy against neutron numbers
at specific proton numbers indicated within the figures.}
\label{fig:1}
\end{figure*}

\begin{figure*}
\includegraphics{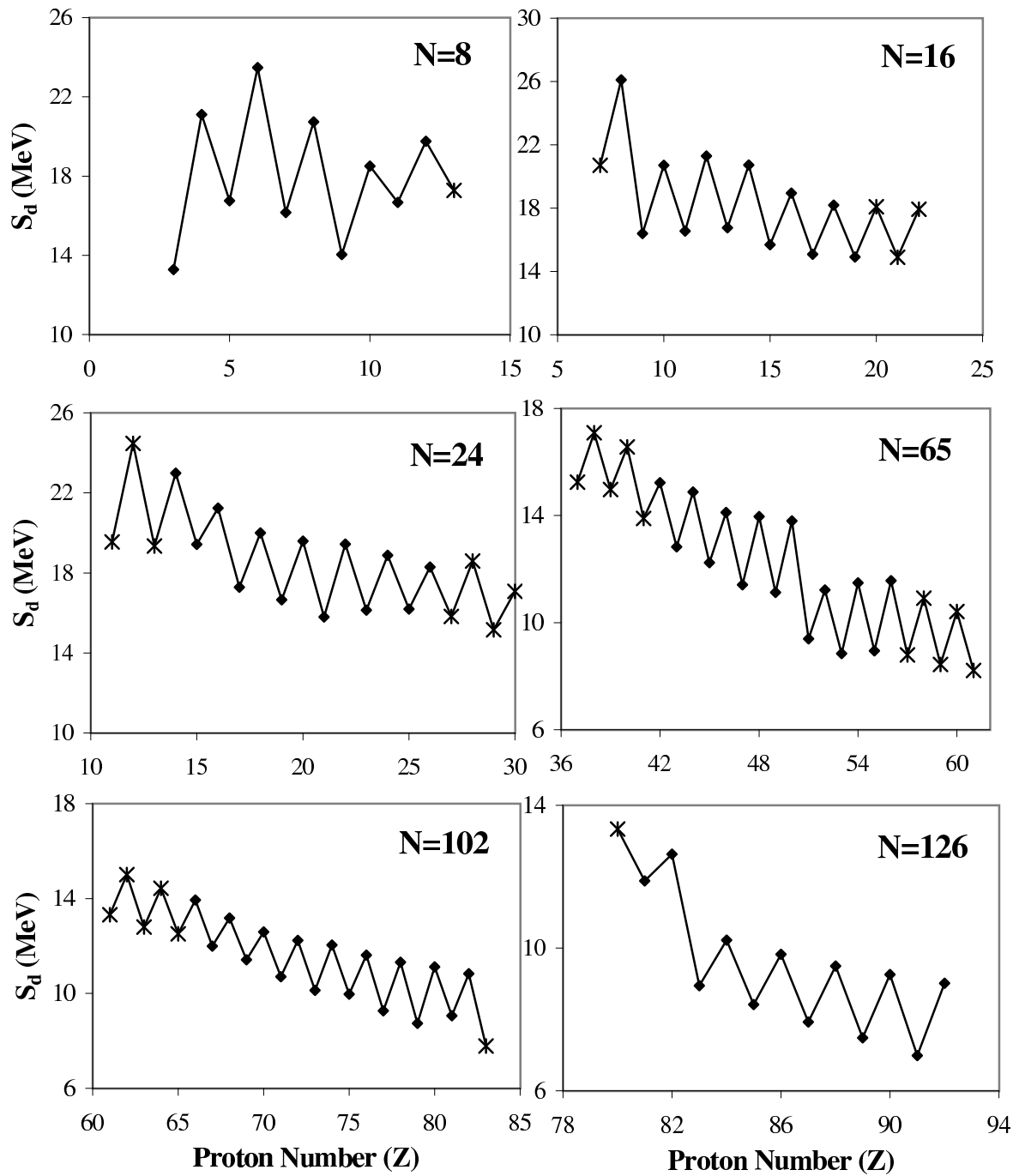}
\caption{Same as Fig.1 except against proton numbers at specific
neutron numbers.}
\label{fig:2}
\end{figure*}

\begin{figure*}
\includegraphics{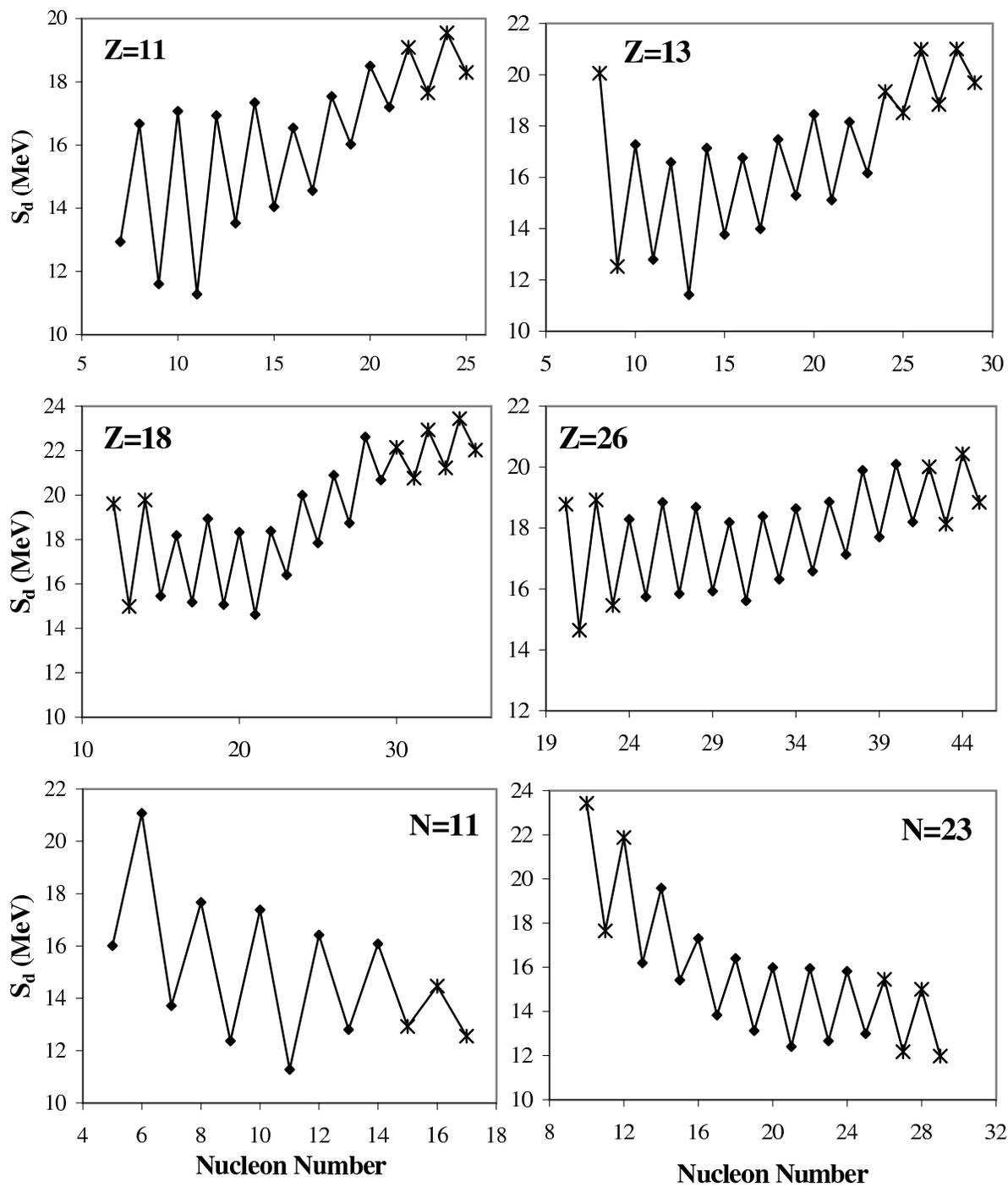}
\caption{Same as Fig.1 except against nucleon numbers (proton or
neutron numbers).}
\label{fig:3}
\end{figure*}

\begin{figure*}
\includegraphics{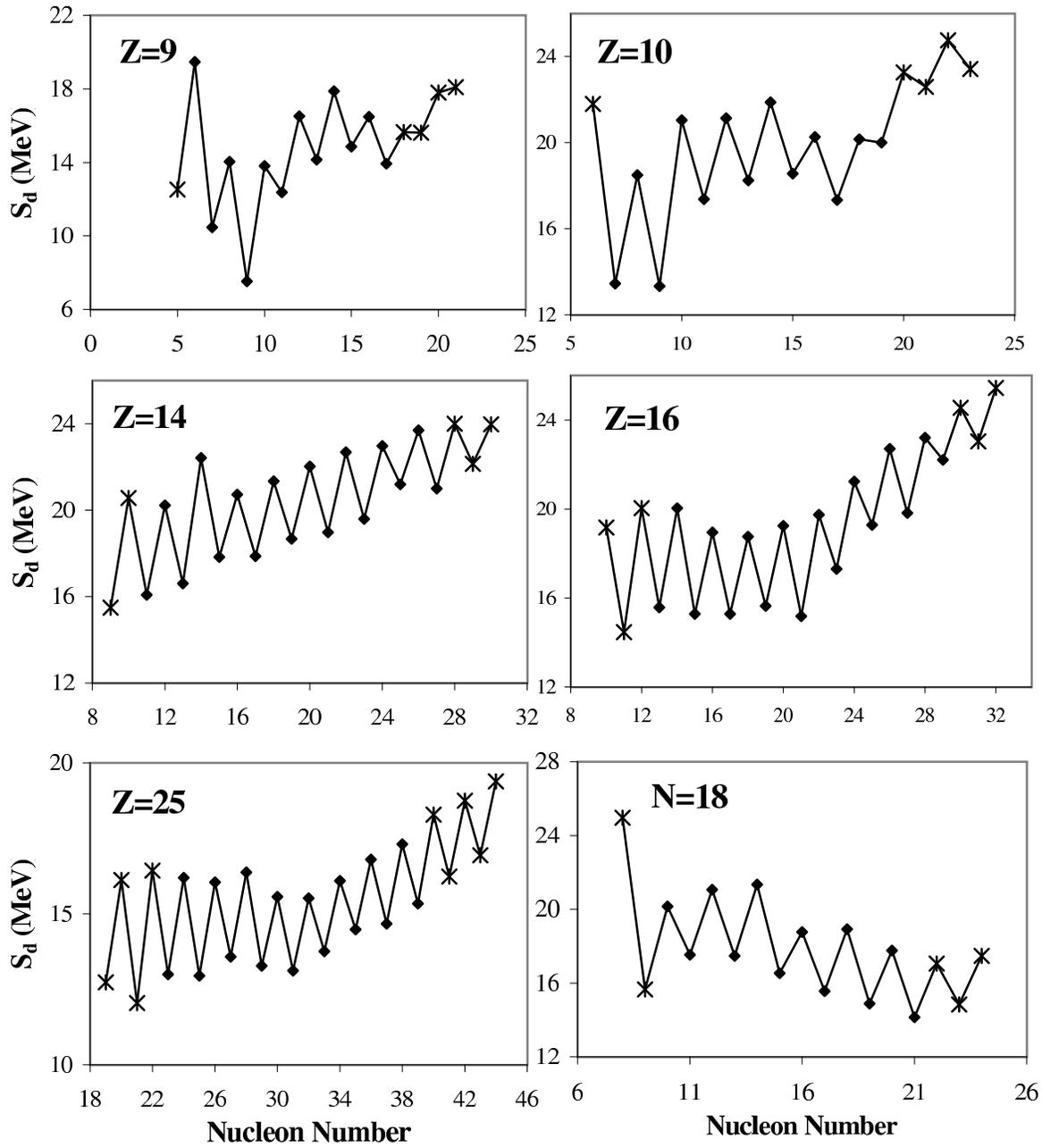}
\caption{Same as Fig.3.}
\label{fig:4}
\end{figure*}

\end{document}